\documentclass[conference]{IEEEtran}
\usepackage[utf8]{inputenc}
\usepackage{amsmath}
\usepackage{amssymb}
\usepackage[usenames,dvipsnames]{xcolor}
\usepackage{tikz}  
\usepackage{graphicx}
\usepackage{fancyhdr}
\usepackage{url}
\usepackage{breakurl}
\usepackage{caption}
\usepackage{subcaption}
\usepackage{tabularx}
\usepackage[noabbrev]{cleveref}
\usepackage{rotating}
\usepackage{multicol}
\usepackage{txfonts}
\usepackage{xcolor}
\newcolumntype{L}[1]{>{\raggedright\arraybackslash}p{#1}} 
\newcolumntype{C}[1]{>{\centering\arraybackslash}p{#1}} 
\newcolumntype{R}[1]{>{\raggedleft\arraybackslash}p{#1}} 

\definecolor{darkspringgreen}{rgb}{0.09, 0.45, 0.27}

\definecolor{chris-comment}{cmyk}{0.0, 0.72, 0.72, 0.28}

\begin{document}

\title{Exploring Security Economics in \\ IoT Standardization Efforts}

\author{\IEEEauthorblockN{Philipp Morgner and Zinaida Benenson}
\IEEEauthorblockA{Friedrich-Alexander-Universität Erlangen-Nürnberg, Germany\\
\{philipp.morgner, zinaida.benenson\}@fau.de}}

\IEEEoverridecommandlockouts
\makeatletter\def\@IEEEpubidpullup{6.5\baselineskip}\makeatother
\IEEEpubid{\parbox{\columnwidth}{
    Workshop on Decentralized IoT Security and Standards (DISS) 2018 \\
    18 February 2018, San Diego, CA, USA \\
    ISBN 1-891562-51-7 \\
    https://dx.doi.org/10.14722/diss.2018.23009 \\
    www.ndss-symposium.org
}
\hspace{\columnsep}\makebox[\columnwidth]{}}

\maketitle

\begin{abstract}
The Internet of Things (IoT) propagates the paradigm of interconnecting billions of heterogeneous devices by various manufacturers. 
To enable IoT applications, the communication between IoT devices follows specifications defined by standard developing organizations. 
In this paper, we present a case study that investigates disclosed insecurities of the popular IoT standard ZigBee, and derive general lessons about security economics in IoT standardization efforts. 
We discuss the motivation of IoT standardization efforts that are primarily driven from an economic perspective, in which large investments in security are not considered necessary since the consumers do not reward them. 
Success at the market is achieved by being quick-to-market, providing functional features and offering easy integration for complementors. 
Nevertheless, manufacturers should not only consider economic reasons but also see their responsibility to protect humans and technological infrastructures from being threatened by insecure IoT products. 
In this context, we propose a number of recommendations to strengthen the security design in future IoT standardization efforts,
ranging from the definition of a precise security model to the enforcement of an update policy.
\end{abstract}


\section{Introduction}
\label{sec:intro}

Standardization efforts play a major role in the expansion of the Internet of Things (IoT) as the success of the IoT is driven by interconnecting a multitude of devices, possibly produced by various manufacturers.
This requires manufacturers to agree on common communication protocols at network and application level.
Such standardization efforts are mainly fostered by standard developing organizations (SDOs). 
In the last years, a number of open and market-driven IoT standardization efforts aimed for market dominance.
In the domain of smart home applications, one of the market leaders is the ZigBee standard, maintained by the ZigBee Alliance, a global non-profit SDO.
In this paper, we consider the ZigBee specifications as a case study to derive lessons about security economics in market-driven IoT standardization efforts.
We chose ZigBee since a number of security weaknesses have been recently revealed in its specifications, and discuss how these weaknesses resulted from design choices made during the standardization process.

To define a suitable security architecture, it is important to understand the motivations of both sides, security researchers and manufacturers, because former research \cite{anderson2001,anderson2006} has concluded that the academic research community has different priorities in securing technologies than the vendors and consumers. 
In fact, the economic perspective is often not considered in the security research. 
Using the lessons from the insecurities of ZigBee as an example, the main goal of this paper is to {raise the understanding on both sides in order to strengthen the security of future IoT standardization efforts}. 

Our contributions are the following: 
We are the first that analyze \emph{root causes} that led to the insufficient security architecture of a popular ZigBee application standard. 
Learning from the security trade-offs made in these IoT specifications, we provide \emph{recommendations} on how to strengthen security architectures in future IoT standardization efforts.
%
Our results show that the majority of the revealed attacks, which ultimately allow the complete take-over of the target devices, could have been prevented if particular compromises in the security design would have been avoided. 
%
We hope that the lessons learned from these security pitfalls raise the attention of manufacturers and SDOs to improve the methodology in defining security measures in future IoT products. 
Also, our goal is that the research community gains a deeper understanding about the economic priorities in IoT standardization efforts.



\section{Background on ZigBee}
\label{sec:background}

ZigBee is an IoT mesh network and application standard maintained by the ZigBee Alliance.
Popular IoT applications that implement the ZigBee standard include smart home products  and smart meters.
The ZigBee Alliance is a non-profit SDO and consists of more than 400 member companies~\cite{zigbee2016}.
The first ZigBee specifications were released in 2004. 
In the early years, the approach of the ZigBee Alliance was to bundle application-specific functionality in separate specifications, denoted as application profiles, to meet the needs of particular applications, e.g., connected lighting systems.
This approach led to problems of interoperability between smart home products that should cooperate in a joint network but implement different application profiles. 
The latest specifications, \emph{ZigBee~3.0}, which were publicly released in 2016, 
aim to unify these profiles into one universal standard. 
In this work, we focus on the ZigBee Light Link specifications for connected lighting.
Although these specifications itself are deprecated since 2016, major parts of them are inherited to ZigBee 3.0.

In ZigBee, each personal area network (PAN) has its own \emph{network key} that is shared among all nodes of this network.
Implementers of products that follow the ZigBee Light Link specifications can choose between two commissioning procedures to obtain the network key:
either classical commissioning or touchlink commissioning.
Classical commissioning is suitable if the network is commissioned using a mobile device application and a bridge device.
In contrast, touchlink commissioning, which was specifically designed for the needs of connected lighting systems, is utilized for managing a network using a constrained device, such as a remote control.  

However, a number of security weaknesses have been revealed in both commissioning modes.
Zillner and Strobl~\cite{zillner2015} demonstrated insecurities in the classical commissioning as they exposed that ZigBee-certified products can be forced to encrypt the network key for the over-the-air transmission using a publicly known fallback key. 
Also, they showed that ZigBee-certified products can be easily reset by sending an unauthenticated reset-to-factory request.

In our previous work~\cite{DBLP:conf/wisec/MorgnerMBMA17}, we analyzed the security of the touchlink commissioning procedure.
We showed that this commissioning procedure is insecure by design, allowing attackers to trigger the identify action (e.g., blinking) of ZigBee-certified devices for several hours, and to change their wireless channel  (and thus permanently disconnect nodes from their legitimate network)  without knowing any key material.
In addition, an attacker can passively eavesdrop the network key and take full control over devices since the master key, which protects the network key during the over-the-air transport, was leaked~\cite{mayazigbee}.
We demonstrated these vulnerabilities by evaluating popular ZigBee-certified connected lighting systems.

In ZigBee 3.0, the mechanisms of the classical commissioning procedure merged with novel features, such as link keys derived from an install code that is printed on the product, into the so-called `EZ-mode' commissioning.
EZ-mode is only activated  for a short period of time after pushing a button on the product.
Thus, the attacks demonstrated by Zillner and Strobl \cite{zillner2015} are mainly contained. 
Also, the ZigBee~3.0 specifications inherited the touchlink commissioning procedure from the ZigBee Light Link specifications with just small adjustments but without replacing the leaked key. 
Thus, the threats shown in our previous work \cite{DBLP:conf/wisec/MorgnerMBMA17} also affect all products certified for ZigBee 3.0 that enable the optional touchlink commissioning.


\section{Root Cause Analysis}
\label{sec:standards} 

The security research community has a different perspective on securing IoT technologies than manufacturers since the economic perspective is often not taken into account.
In this section, we outline priorities and incentives of market-driven standardization efforts and then analyze what went wrong in the standardization of the ZigBee Light Link specifications.\footnote{The authors are not associated with the ZigBee Alliance. The information presented in this case study were obtained by discussions with officials and members of the ZigBee Alliance, publicly accessible information including specifications, and technical inspections of ZigBee-certified products.}

\subsection{Motivation for Standardization}

Several analyses (e.g., \cite{blind, hagedoorn1993, glaister1996}) outline motivations of manufacturers to participate in strategic alliances, which promote the standardization of novel technologies.

A first reason for participation is to decrease market uncertainties since the risks are shared among all participating companies \cite{swann2000, hagedoorn1993}. 
For the innovator, standardization increases the probability that the own technology succeeds, and it prevents other alliance members from developing competitive (proprietary) systems. 
In the case of the ZigBee Light Link specifications that define a network and application standard for connected lighting, the ZigBee Alliance started the development of the standard in 2010, with contributions from Philips, Osram, and GE, among others. 
 The ZigBee Light Link technology became a large success since another organization, The Connected Lighting Alliance (TCLA), a non-profit organization  promoting the compatibility of wireless lighting, endorsed this standard in July 2013 after studying multiple open standards. 

A second reason is that members of standardization processes profit from strategic knowledge transfer among alliance members \cite{buckley2009}. 
Since multiple manufacturers contribute their know-how to the standardization efforts, alliance members benefit from knowledge spillover, as well as keep track over technical knowledge of their potential competitors \cite{sampson2007}.  
According to the ZigBee Alliance, the contribution of intellectual property to their standards by member companies is very common. 
As an example of knowledge transfer, the touchlink commissioning procedure, intellectual property of Philips \cite{philips:patent}, was contributed to the ZigBee Light Link specifications. 

Access to new markets is a third reason for participation in standardization efforts. Alliances provide low entry levels for entering foreign markets, i.e., markets that have not been entered by a manufacturer yet \cite{glaister1996}. 
Also making own products compatible to complementary products opens new markets, even for small companies.
The ZigBee Light Link standard was developed because members of the ZigBee Alliance saw a promising market. 
Afterwards, further companies that did not participate in the development of these specifications, offered products that complement ZigBee-certified lighting systems, e.g., wireless dimmer switches  \cite{busch-jaeger}.

To bear the expenses of the organizational overhead of such a strategic alliance, their members are obligated to pay an annual fee.
Usually, SDOs offer several levels of membership that differ in the amount of  fee and privileges: the more financial resources a member contributes to the alliance, the more influence this member has on the alliance's final decisions.
In case of the ZigBee Alliance, three membership levels are offered: adopter (\$4k/year), participant (\$9.9k/year), and promoter (\$55k/year) \cite{zigbeemember}.
While adopters have access to all final specifications and some group events, only participants and promoters can participate in work groups and propose specifications.
Of them, only promoters have the right to finally approve new specifications.

\subsection{ZigBee as Case Study on Security Economics}

From the economic perspective \cite{anderson2001, anderson2006}, if a standard is aiming for market dominance, then this standard must attract manufacturers of complementary products as well as consumers. These prioritized efforts take much resources, and since resources are finite, they tend to be withdrawn from non-functional features, e.g., comprehensive security measures. 
At the end, a large amount of resources is spent to develop an attractive system but only a few resources are left to make it secure.
In fact, security measures may even make it harder for complementors to build complementary products that support this standard. 
Therefore, in the first phase of an evolving technology, manufacturers tend to ignore security as they expand their market position.
Consumers reward manufacturers for adding functional features to products and being first at the market. 
On the contrary, the development of an adequate security architecture for these products requires time-consuming testing and might restrict favored functional features. 
In a latter phase, security measures may be added to lock consumers to the products. 
These two phases can be seen in the ZigBee Light Link standard.

\subsubsection*{Phase 1 -- Security Design Trade-Offs} 
The ZigBee specifications prior to ZigBee~3.0 distinguished between home consumer and business applications. 
In the case of connected lighting systems, the ZigBee Light Link standard was intended to serve the home consumer market, while another ZigBee application profile, the ZigBee Building Automation standard~\cite{spec:zigbee:ba}, provides functionality for connected lighting systems in business and industrial settings.
The significant differences between these two standards can be found in their security architectures. While ZigBee Building Automation follows the ZigBee Pro specifications and offers the \emph{full} classical commissioning procedure that uses a dedicated device, called Trust Center, for key management \cite[p.432]{spec:zigbee:pro}, the ZigBee Light Link standard aimed to decrease the complexity of the commissioning in order to increase consumer acceptance. 
Thus, the classical commissioning procedure in the ZigBee Light Link standard lacks the  Trust Center and relies on a global non-disclosure agreement (NDA)-protected master key.

Attacks against the classical commissioning procedure are known. 
These attacks exploit fallback mechanisms that are in place to compensate the lack of the Trust Center \cite{zillner2015}.
Security weaknesses were also found in the second commissioning mode, the touchlink commissioning procedure. 
Touchlink commissioning uses the \emph{inter-PAN transmission} mechanism to join a new device to an existing network, one of the most security-critical operations in ZigBee networks including the transport of the network key to the joining device.
The concept of inter-PAN frames was adopted from the then already existing ZigBee Smart Energy specifications, a profile targeting smart metering applications. 
The ZigBee Smart Energy specifications define the purpose of inter-PAN transmissions as possibility for ZigBee devices to `\emph{perform limited, insecure, and possibly anonymous exchange of information}' \cite[p.81]{spec:zigbee:se}. 
An exemplary application utilizes this transmission mechanism for the `\emph{market requirement to send pricing information to very low cost devices}', e.g., a refrigerator magnet that displays the current energy prices and consumption.
The ZigBee Light Link specifications adopted these inter-PAN transmission mechanism to enable the commissioning of networks with constrained devices.
Intended use cases are, e.g., a bulb that should be joined to an existing network using a simple remote control.
The adoption of these unauthenticated inter-PAN transmissions to reduce the complexity of commissioning procedures, in combination with the usage of signal strength as physical security measure, resulted in the insecurities presented in \cite{DBLP:conf/wisec/MorgnerMBMA17}.

Another critical point is the trust in the safe-keeping of master keys that are shared among multiple manufacturers.
Both commissioning procedures of the ZigBee Light Link standard rely on an NDA-protected shared key used to encrypt the network key. 
Although the \emph{distributed security global link key} (also known as \emph{ZLL link key}), used for the classical commissioning (and its successor in ZigBee 3.0: EZ-mode commissioning), is not leaked yet, this can happen anytime. 
NDA-protected keys can indeed leak as demonstrated by the \emph{touchlink preconfigured link key} (also known as \emph{ZLL master key}), which was leaked in March 2015 on Twitter \cite{mayazigbee}. 

All these security weaknesses, resulting from over-simplified (thus insecure) commissioning procedures including master keys and fallback mechanisms, show that trade-offs have been made at the expenses of a comprehensive security architecture to allow other manufacturers to adopt to this standard easily.
At the same time, ZigBee-certified products implementing the more secure ZigBee Building Automation standard have not been released yet. 
The separation between home consumer and business applications has been discontinued in ZigBee~3.0, while the touchlink commissioning procedure is still an optional feature in ZigBee~3.0.

\subsubsection*{Phase 2 -- Lock the Consumers} 
In December 2015, Philips (as one of the driving forces behind the ZigBee Light Link standard) assumingly tried to lock consumers more tightly to its products. 
This happened with an update of the Hue app, which locked out products from other vendors like Osram and GE that are not participating in the `Friends of Hue' certification program. 
The public sentiment was large such that Philips reverted this decision after a few days through providing a non-scheduled update \cite{hue-friend}. 
Although Philips did not disclose the mechanism how the lock-out was technically implemented, this mechanism can be seen as a security feature that restricts access to the network for white-listed devices.

\subsubsection*{Consumers' Difficulties with Assessing Security}
As described in the economic theory of `the market for lemons'~\cite{akerlof}, consumers are unwilling to pay for something they cannot assess, such as security \cite{anderson2001}. 
The ZigBee specifications define security measures that are partly very ineffective. 
But how can a regular consumer determine which level of security is provided by an IoT product?
The security of the ZigBee specifications seem solid at first sight as the specifications apply the well-known encryption and authentication scheme AES-CCM. 
Assuming a shared network key, an attacker without knowledge of this key is not able to decrypt or manipulate AES-CCM-encrypted messages.
Nevertheless, as shown in \cite{zillner2015, DBLP:conf/sp/RonenSWO17, DBLP:conf/wisec/MorgnerMBMA17}, an attacker is able to gain full control of ZigBee-certified products.


\section{Implications of Insecure IoT Products}
\label{sec:benefits} 

So far, it seems that the development of strong security measures in IoT products mainly leads to competitive disadvantages.
From an economic perspective, large investments in security are not necessary as the consumers do not reward them. 
High priorities are being quick-to-market, providing functional features and offering easy integration for complementors.
One might assume that investments in security saves money in the long run: 
Through security breaches, people would lose  their trust in the  manufacturer and the companies' reputation decreases. 
But experiences from the past show that many companies, which have been affected by trust-losing data breaches, do not go out of the business, although they may suffer significant short-term consequences~\cite{DBLP:journals/ijecommerce/CavusogluMR04, acquisti2006there, DBLP:journals/jcs/GordonLZ11, lange2017}. 
For example, security flaws in connected lighting systems (see \cite{zillner2015, DBLP:conf/sp/RonenSWO17, DBLP:conf/wisec/MorgnerMBMA17,dhanjani2013, chapman2014, rapid7}) have been disclosed almost since the release of these products but none of them really affected the attraction in this technology or its vendors.
From the economic perspective of the manufacturers, there are less benefits in strongly securing IoT devices compared to the benefits that arise from shorter development cycles omitting these security measures. 
From our point of view, this might change in the future, when people start suing manufacturers of insecure IoT products for financial compensation, or governmental regulations demand comprehensive security levels for market entry.

Irrespective of the current situation, we state that manufacturers should take into account ethical considerations and act responsibly.
Consumers can be indeed harmed through the insecurities of IoT systems, even by connected lighting systems. 
Blackouts or unintended blinking of lights may not only annoy but also frighten residents. 
Epileptic seizures can be caused by flickering lights on epilepsy patients~\cite{DBLP:conf/eurosp/RonenS16, wilkins2010}.
Insecure IoT systems have been used for large denial-of-service attacks against critical infrastructures~\cite{krebs2016, DBLP:conf/uss/AntonakakisABBB17}. 
Researchers showed that even a single infected light bulb has the potential to serve as incubator to spread malware on IoT systems across large areas \cite{DBLP:conf/sp/RonenSWO17}.
All these threats might endanger humans as well as infrastructure.
If manufacturers deny their responsibility to protect against such threats, which clearly result from security weaknesses in their products, they might become subjects to class action lawsuits. 
Recently, there have been several class action lawsuits against manufacturers that acted irresponsibly. 
Volkswagen settled a class action lawsuit on its diesel emissions cheating scandal by paying compensations of 14.7 billion USD \cite{vw-scandal}.
Samsung faced class action lawsuits for the slow replacement of its fire-prone mobile phone model \cite{samsung-scandal}.
In another class action lawsuit, Ford is alleged of knowingly releasing a flawed infotainment system \cite{ford-scandal}.
These lawsuits have high economic impacts on manufacturers and could have been avoided by taking responsibility seriously.


\section{A Road to Improvement}
\label{sec:recommendation} 

We propose five recommendations to strengthen the security design of future IoT standardization efforts. 

\subsection{Define Precise Security Models}
\label{sec:recommendation:model}
During our research, we realized that most specifications of IoT standards do not define a precise security model.
The objective of the security model is to formulate against what threats an IoT system should be protected (`security goals') and who are the potential attackers (`attacker model' or `threat model'). 
In fact, this model formulates the goals of the security design, and therefore, should also be part of the specification.
Based on the security model, the security architecture should be developed  that considers potential threats comprehensively.
Such an architecture provides a significantly smaller attack surface than security architectures designed by experience (or best practices) but without assessing the specific threat conditions of this  application.
As V. D. Gligor said referring to security models for wireless ad-hoc networks: 
``A system without an adversary definition cannot possibly be insecure; it can only be astonishing, and of course astonishment is a much underrated security vice'' \cite{DBLP:conf/spw/Gligor08}. 
Exemplary, we reviewed the ZigBee Light Link specifications~\cite{spec:zigbee:ll}, the ZigBee 3.0 base device behavior and cluster library specifications \cite{spec:zigbee3:base, spec:zigbee3:cluster}, but also the LoRaWAN \cite{spec:lorawan} and the  Bluetooth 5.0  specifications \cite{bluetooth5} regarding security models.  
These IoT-related standards are widely supported by large alliances and their specifications are released to the public.
All these standards lack the definitions of an attacker model as well as the security goals. 
Although this recommendation demands more extensive periods of standard development cycles, the process of developing a security model can be designed in a generic way, such that the resulting model can be adopted to different applications with small effort.
The development of the ZigBee Light Link standard took more than two years. Compared to this period of time, the discussion and definition of a comprehensive security model should not increase this duration significantly. 

\subsection{Stop Consumer and Business Security Differentiation}
As described in \Cref{sec:standards}, some SDOs tend to distinguish between home consumer and business products. 
In the ZigBee Light Link standard, the security of home consumer products is based on a weaker security architecture than business products of the comparable ZigBee Building Automation standard.
If we  compare the volume of sold products, then more IoT home consumer products than business IoT products are sold. Gartner predicted the installation of around 7 billion consumer  IoT devices compared to 4.2 billion business IoT devices  in 2018 \cite{gartner2017}.
Thus, a security breach of a popular IoT home consumer products would affect millions of devices.
However, there is no straight line between consumer and business products in IoT technologies. 
Although the Philips Hue system is intended for home consumer use, it can also be deployed in an industrial context, e.g., to control workflows~\cite{fortune}.
Since many IoT systems offer interfaces for third-party applications, the implementation of consumer products in industrial processes is a simple and inexpensive option. 
In the case of connected lighting systems based on ZigBee Light Link, another reason might be that there exists no business product (based on the ZigBee Building Automation standard) that offers similar functionalities and flexibility to the best of our knowledge. 
Thus, we state that the IoT demands high security standards for both, consumer and business products.

\subsection{Add Membership Level for Academic Institutes}
We counted the contribution of academic researchers to popular IoT-related standards and assessed exemplary the IoT standards mentioned in \Cref{sec:recommendation:model}. 
In none of these specifications, academic institutes are listed as contributors, except Bluetooth 5.0 states contributions by the University of Bonn (and NIST).
Manufacturers have extensive experience in the aspects of functionality and the needs of the market, which are very important insights. 
Moreover, we assume that they employ experienced security engineers. 
However, corporate security engineers might be too much aware of the business goal trade-offs involved in the security design.
This conflict of interest may make it difficult for them to insist upon meeting strong security goals. 
Therefore, an outside view of the academic researchers can help in two ways: (1) to better appraise the probabilities of attacks and consequences of insecure design and (2) to integrate innovative research solutions into the security design.
To achieve academic participation, SDOs should lower the barriers for contribution, e.g., by introducing a membership level for (selected) academic institutes that allows (and potentially pays) academic experts to participate in work groups. 
While fostering academic participation requires investments in selecting and hiring academic experts, the opportunity of knowledge spillover from academic research can be a valuable enrichment to standardization efforts.

\subsection{Conduct Security Testing Without Conflict of Interest}
During the final certification process of the product, not only the functionality of the product should be tested for compliance. 
Also the implementation of the security architecture in software and hardware should be evaluated with code audits and security-focused penetration tests by external certification labs.
Certain attacks \cite{DBLP:conf/sp/RonenSWO17, DBLP:conf/wisec/MorgnerMBMA17} on connected lighting systems exploited implementation bugs that could have been avoided through external security testing. 
However, security testing has potential points of failure:
Previous investigations \cite{DBLP:journals/ieeesp/MurdochBA12, leverett2017standardisation} showed that vendors often prefer testing and certification labs that perform a relaxed evaluation.
Also, testing labs might lose customers if they are too strict and delay the release of products since their competitors might be more easygoing.
Thus, mechanisms must be in place that precisely define the scope of evaluation, the exact testing procedures, and punishment for certification labs that fail to fulfill these requirements. 
To put the SDO in charge of supervising the certification labs leads to a conflict of interest.
The SDO's economic interest is to bring products fast to the market to gain market dominance, which corrupts the motivation for a comprehensive and time-consuming security testing.
Therefore, an independent entity that is not influenced by economic motivations must be in charge to supervise the certification labs.
The conduction of these security penetration tests will most likely increase costs and demands more extensive periods of product development cycles but finally also reduces the probability of expensive replacement and patching of installed devices.

\subsection{Define and Enforce Update Policy}
During the standardization process, not only security mechanisms against currently known attacks should be considered but also the possibility of upgrading security mechanisms in case novel attacks are disclosed, implementation bugs are discovered, or more efficient security measures have been found. 
While most current IoT standardization efforts consider update mechanisms, an update policy is usually not defined by the SDO.
The update policy defines under what circumstances a product needs to be updated and within which time frame.
In addition, this policy should define who takes responsibility for updates if the vendor is not able to deliver them anymore.
If such an update policy is in place (and executed), then the security of the IoT application should be ensured for its full lifetime.
The problem of such an update policy is the enforcement: what should be the motivation of the manufacturer to update its legacy products?
From the economic perspective, why should companies invest money if there is no profit?
Thus, the motivation must be extrinsic.  
A regulatory paradigm shift might be necessary such that only lifetime security-providing vendors gain access to the markets.
For instance, manufacturers that fail to fulfill their duties in terms of providing updates for their products would not be allowed to receive the certification for the release of new products.
The SDO cannot be trusted with the enforcement of such an update policy due to its conflict of interest.
Hence, an independent entity, which is not driven by economic motivations, is required to enforce this policy.


\section{Related Work}
\label{sec:related}

Research on security economics in IoT and standardization gained little attention so far.
In terms of distributed systems, Anderson and Fuloria \cite{DBLP:conf/weis/AndersonF10} investigated the security economics of electricity metering.
%
Murdoch et al.\ \cite{DBLP:journals/ieeesp/MurdochBA12} analyzed reasons why certified products fail to fulfill standardized security requirements. 
Levä et al.\ \cite{DBLP:journals/csi/LevaS13} proposed a framework to analyze the economic feasibility of protocols during standard development.
Ray et al.\ \cite{DBLP:conf/iccd/RayHBB16} outlined trade-offs between energy and security constraints in the IoT ecosystem.
Leverett et~al.~\cite{leverett2017standardisation} described problems and opportunities regarding the standardization and certifications of IoT applications in terms of safety, security, and privacy.
Focusing on the European Union, they proposed to establish institutional resources for regulators and policy-makers. 
In contrast to our approach, Leverett et al.\  proposed actions at state-level, while our work investigates the security economics of IoT standardization efforts in the private sector.
%
To the best of our knowledge, we are the first that investigate root causes of insecurities in a specific IoT standard.
Learning from them, we deduct lessons about security economics in IoT standards and recommend principles to improve the outcome of future IoT standardization efforts.


\section{Conclusion}
\label{sec:conclusion}

In the past years, security weaknesses were disclosed in ZigBee specifications, one of the most popular IoT standards in the domain of smart homes:  from leaked master keys and fallback mechanisms to unauthenticated command messages.
Similar flaws are not only specific to ZigBee but also in other IoT standards, e.g., Bluetooth Low Energy \cite{DBLP:conf/woot/Ryan13}.
Learning from the security pitfalls of the ZigBee specifications, we analyzed the root causes for these insecurities and found them in the prioritization of market aspects over  a comprehensive security design.
More focus on designing security measures during  IoT standardization efforts is needed to protect against the rising threats that result from billions of interconnected IoT devices.


\section*{Acknowledgments}

This work is supported by the German Research Foundation (DFG) under Grant AR 671/3-1.
We thank Felix Freiling, Gaston Pugliese, and the anonymous reviewers for helpful comments.


\bibliographystyle{IEEEtran}
\bibliography{bib-zll-philipp,bib-z3,bib-econ}

\end{document}